\documentclass[12pt,prd,aps,amssymb,amsmath,tightenlines,showpacs,a4paper,nofootinbib]{revtex4}

\def\bea{\begin{eqnarray}}
\def\eea{\end{eqnarray}}
\def\half{\frac{1}{2}}
\def\vf{\varphi}

\begin{document}
\title{The Floquet Method for $PT$-symmetric Periodic Potentials}

\author{H.~F.~Jones\email{h.f.jones@imperial.ac.uk}}

\affiliation{
Physics Department, Imperial College, London SW7 2BZ, UK}
\date{\today}

\begin{abstract}
By the general theory of $PT$-symmetric quantum systems, their energy levels are either real or occur in complex-conjugate pairs, which
implies that the secular equation must be real. However, for periodic potentials it is by no means clear that the secular equation
arising in the Floquet method is indeed real, since it involves two linearly independent solutions of the Schr\"odinger equation. In this brief note we elucidate how that reality can be established.
\end{abstract}


\maketitle

The study of quantum mechanical Hamiltonians that are $PT$-symmetric but not Hermitian\cite{BB}-\cite{AMh} has recently found an unexpected application in classical optics\cite{op1}-\cite{op6}, due to the fact that in the paraxial approximation the equation of propagation of an electromagnetic wave in a medium is formally identical to the Schr\"odinger equation, but with different interpretations for the symbols appearing therein. The equation of propagation takes the form
\bea\label{opteq}
i\frac{\partial\psi}{\partial z}=-\left(\frac{\partial^2}{{\partial x}^2}+V(x)\right)\psi,
\eea
where $\psi(x,z)$ represents the envelope function of the amplitude of the electric field, $z$ is a scaled propagation distance, and $V(x)$ is the optical potential, proportional to the variation in the refractive index of the material through which the wave is passing. That is, $V(x)\propto v(x)$, where $n=n_0(1+v(x))$, with $|v|\ll 1$. A complex $v$ corresponds to a complex refractive index, whose imaginary part represents either loss or gain. In principle the loss and gain regions can be carefully configured so that $v$ is $PT$ symmetric, that is $v^*(x)=v(-x)$.

In the optical context one is often interested in periodic $PT$-symmetric potentials, which have a number of very interesting properties\cite{op1}-\cite{op6}. Such a potential $V(x)$,
whose period we can take as $\pi$, without loss of generality, satisfies the two conditions
$V^*(-x)=V(x)=V(x+\pi)$. For a periodic potential we are interested in finding the Bloch solutions, which are solutions of the
time-independent Schr\"odinger equation
\bea\label{SE}
-\left(\frac{\partial^2}{{\partial x}^2}+V(x)\right)\psi_k(x)=E\psi_k(x)
\eea
with the periodicity property $\psi_k(x+\pi)=e^{ik\pi}\psi_k(x)$.

The Floquet method consists of expressing $\psi_k(x)$ in terms of two solutions, $u_1(x)$ and $u_2(x)$, of Eq.~(\ref{SE}), with initial conditions
\bea
u_1(0)=1, &\phantom{=}&  u'_1(0)=0 \ , \cr
u_2(0)=0, &\phantom{=}&  u'_2(0)=1 \ .
\eea
Then $\psi_k(x)$ is written as the superposition
\bea
\psi_k(x)=c_k u_1(x) +d_k u_2(x).
\eea
Imposing the conditions $\psi_k(\pi)=e^{ik\pi}\psi(0)$ and $\psi'_k(\pi)=e^{ik\pi}\psi'(0)$ and exploiting the invariance of the Wronskian
$W(u_1,u_2)$ one arrives at the secular equation
\bea
\cos{k\pi}=\Delta \equiv \half\left(u_1(\pi)+u'_2(\pi)\right) \ .
\eea
In the Hermitian situation both $u_1(\pi)$ and $u_2(\pi)$ are real, and the equation for $k$ gives real solutions (bands) when
$|\Delta|\le 1$. However, in the non-Hermitian, $PT$-symmetric, situation it is not at all obvious that $\Delta$ is real, since that
implies a relation between $u_1(\pi)$ and $u_2'(\pi)$, even though $u_1(x)$ and $u_2(x)$ are linearly independent solutions of Eq.~(\ref{SE}). It is that problem that we wish to address in the present note. In fact we will show that $u'_2(\pi)=u^*_1(\pi)$.

The clue to relating $u_1(\pi)$ and $u_2(\pi)$ comes from considering a half-period shift, namely $x=z+\pi/2$. We write $\vf(z)=\psi(z+\pi/2)$ and
$U(z)=V(z+\pi/2)$. Then $\vf(z)$ satisfies the Schr\"odinger equation
\bea\label{NSE}
-\left(\frac{\partial^2}{{\partial z}^2}+U(z)\right)\vf_k(z)=E\psi_k(z)\ .
\eea
The crucial point is that because of the periodicity and $PT$-symmetry of $V(x)$ the new potential $U(z)$ is also $PT$-symmetric. Thus $U(-z)=V(-z+\pi/2)=V(-z-\pi/2)=V^*(z+\pi/2)=U^*(z)$.

Now we can express the Floquet functions $u_1(x)$, $u_2(x)$ in terms of Floquet functions $v_1(z)$, $v_2(z)$ of the transformed equation (\ref{NSE}), satisfying
\bea
v_1(0)=1, &\phantom{=}&  v'_1(0)=0 \ , \cr
v_2(0)=0, &\phantom{=}&  v'_2(0)=1 \ .
\eea
It is easily seen that the relation is
\bea\label{shift}
u_1(x)&=&\phantom{-}v_2'(-\pi/2)v_1(z)-v_1'(-\pi/2)v_2(z)\cr
u_2(x)&=&-v_2(-\pi/2)v_1(z)+v_1(-\pi/2)v_2(z) ,
\eea
in order to satisfy the initial conditions on $u_1(x)$, $u_2(x)$.\\
So
\bea
u_1(\pi)&=&\phantom{-}v_2'(-\pi/2)v_1(\pi/2)-v_1'(-\pi/2)v_2(\pi/2)\cr
u_1'(\pi)&=&\phantom{-}v_2'(-\pi/2)v_1'(\pi/2)-v_1'(-\pi/2)v_2'(\pi/2)\cr
u_2(\pi)&=&-v_2(-\pi/2)v_1(\pi/2)+v_1(-\pi/2)v_2(\pi/2)\cr
u_2'(\pi)&=&-v_2(-\pi/2)v_1'(\pi/2)+v_1(-\pi/2)v_2'(\pi/2)\ .
\eea
But, because of $PT$-symmetry,
\bea
v_1(-\pi/2)&=&\phantom{-}(v_1(\pi/2))^*\cr
v_1'(-\pi/2)&=&-(v_1'(\pi/2))^*\cr
v_2(-\pi/2)&=&-(v_2(\pi/2))^*\cr
v_2'(-\pi/2)&=&\phantom{-}(v_2'(\pi/2))^*\ .
\eea
Hence, indeed, $u_1(\pi)=(u_2'(\pi))^*$, and moreover, $u_1'(\pi)$ and $u_2(\pi)$ are real.
The statement $u_1(\pi)=(u_2'(\pi))^*$ is in fact the $PT$-generalization of the relation $u_1(\pi)=u'_2(\pi)$
implied without proof by Eq.~(20.3.10) of Ref.~\cite{AS} for the Hermitian case of the Mathieu equation, where $V(x)=\cos(2x)$.

If we wish, we may express everything in terms of $u_1$, $u_2$ because from Eq.~(\ref{shift})
\bea
u_1(\pi/2)&=&\phantom{-}v_2'(-\pi/2)\cr
u_1'(\pi/2)&=&-v_1'(-\pi/2)\cr
u_2(\pi/2)&=&-v_2(-\pi/2)\cr
u_2'(\pi/2)&=&\phantom{-}v_1(-\pi/2)\ .
\eea
Hence
\bea
u_1(\pi)=(u_2'(\pi))^*=u_1(\pi/2)(u_2'(\pi/2))^*+u_1'(\pi/2)(u_2(\pi/2))^*\ ,
\eea
which is the $PT$-generalization of a relation implied by Eq.~(20.3.11) of Ref.~\cite{AS} after the use of the invariance of the Wronskian.

Similarly
\bea
u_1'(\pi)&=&2\mbox{Re}\left(u_1^*(\pi/2)u_1'(\pi/2)\right)\ ,\cr
u_2(\pi)&=&2\mbox{Re}\left(u_2^*(\pi/2)u_2'(\pi/2)\right)\ .
\eea



\begin{thebibliography}{99}
\bibitem{BB} C.~M.~Bender and S.~Boettcher, Phys.~Rev.~Lett. {\bf 80},
5243 (1998).
\bibitem{CMBR} C.~M.~Bender, Contemp.~Phys. {\bf 46}, 277 (2005);
Rep.~Prog.~Phys. {\bf 70}, 947 (2007).
\bibitem{AMR} A.~Mostafazadeh, Int. J. Geom. Meth. Mod. Phys. {\bf 7}, 1191 (2010).
\bibitem{BBJC} C.~M.~Bender, D.~C.~Brody and H.~F.~Jones,
Phys.~Rev.~Lett. {\bf 89}, 270401 (2002) ; 92, 119902(E) (2004).
\bibitem{BBJQ} C.~M.~Bender, D.~C.~Brody and H.~F.~Jones,
Phys.~Rev.~D {\bf 70}, 025001 (2004) ; 71, 049901(E) (2005).
\bibitem{AMh} A.~Mostafazadeh, J.~Math.~Phys. {\bf 43}, 205 (2002);
J.~Phys.~A {\bf 36}, 7081 (2003).
\bibitem{op1} R.~El-Ganainy, K.~G.~Makris, D.~N.~Christodoulides and Z.~H.~Musslimani, Optics Letters {\bf 32}, 2632 (2007).
\bibitem{op2} Z.~Musslimani, K.~G.~Makris, R.~El-Ganainy and D.~N.~Christodoulides,  Phys.~Rev.~Lett. {\bf 100}, 030402 (2008).
\bibitem{op3} K.~Makris, R.~El-Ganainy, D.~N.~Christodoulides and Z.~Musslimani, Phys.~Rev.~Lett. {\bf 100}, 103904 (2008).
\bibitem{op4} K.~Makris, R.~El-Ganainy, D.~N.~Christodoulides and Z.~Musslimani, Phys.~Rev.~A {\bf 81}, 063807 (2010).
\bibitem{op5} S.~Longhi, Phys.~Rev.~A {\bf 81}, 022102 (2010).
\bibitem{op6} E.~M.~Graefe and H.~F.~Jones, Phys.~Rev.~A {\bf 84}, 013818 (2011).
\bibitem{AS} M.~Abramowitz and I.~A.~Stegun, {\sl Handbook of Mathematical Functions}, Dover, NY, 1970.
\end{thebibliography}
\end{document}